# Central indexes to the citation distribution: A complement to the h-index [1]

PABLO DORTA-GONZÁLEZ*

MARÍA-ISABEL DORTA-GONZÁLEZ†

* Departamento de Métodos Cuantitativos en Economía y Gestión
Universidad de Las Palmas de Gran Canaria,
Gran Canaria, España
pdorta@dmc.ulpgc.es

† Departamento de Estadística, Investigación Operativa y Computación
Universidad de La Laguna
Tenerife, España
isadorta@ull.es

[1] This research was partially financed by the Ministerio de Ciencia e Innovación, grant ECO2008-05589.

# Central indexes to the citation distribution: A complement to the h-index

**ABSTRACT**

The citation distribution of a researcher shows the impact of their production and determines the success of their scientific career. However, its application in scientific evaluation is difficult due to the bi-dimensional character of the distribution. Some bibliometric indexes that try to synthesize in a numerical value the principal characteristics of this distribution have been proposed recently. In contrast with other bibliometric measures, the biases that the distribution tails provoke, are reduced by the h-index. However, some limitations in the discrimination among researchers with different publication habits are presented in this index. This index penalizes selective researchers, distinguished by the large number of citations received, as compared to large producers. In this work, two original sets of indexes, the central area indexes and the central interval indexes, that complement the h-index to include the central shape of the citation distribution, are proposed and compared.

## 1. Introduction

There exists a general consensus among researchers about journal articles which are the most direct results of research, especially when journals with a selective process that guarantees quality and originality are considered. Although expert opinion is believed to be the most appropriate method of valuing the contribution of an article to a specific field of knowledge, this system presents some limitations, such as the subjective character and its high cost. In this context, bibliometric indexes represent objective evidences that can be used to complement expert opinion.

It is known that some works of limited success are published by the best journals, and some works of great success are published in journals that are not top ranked according to the impact factor. Therefore, there is some rejection to evaluating the impact of a work by the impact factor of the publishing journal.

Most common indexes used to evaluate researchers are based on counting publications and received citations. The *number of publications* $(N_p)$ is a quantitative indicator that does not value the scientific advance of the contribution. As qualitative indicators able to assess the impact, influence or visibility of a research, the *total number of citations* $(N_c)$ and the *average citations per article* $(n_c = N_c / N_p)$ are used. However, although these indicators show the success of a scientific career in many cases, sometimes isolated successes accumulate a high percentage of the total number of citations. In addition, important biases are introduced by large collaborations that collect many citations derived from the work of a large number of researchers.

The $h$-index [HIRSCH, 2005] tries to solve these limitations. A researcher has an $h$-index when $h$ of its publications have each received at least $h$ citations, and the rest have $h$ or less citations. The number of important articles pertaining to a researcher is estimated by this index, increasing their requirements at the same time as their value rises. Moreover, a lower bound $h^2$ of $N_c$ is provided. The value $N_c$ is generally much greater than $h^2$ (HIRSCH [2005] has estimated between 3 and 5 times greater). This amount underestimates the citations of the $h$ most cited articles (Hirsch core) and ignores articles with less than $h$ citations. A correlation between the $h$-index and the success of a researcher appreciated by his peers has been obtained [HIRSCH, 2005], and the future success of a researcher could be predicted by this value [HIRSCH, 2007].

The $h$-index has been extensively studied (see reviews by BORNMANN & DANIEL [2007]; ALONSO ET AL. [2009]; and the stochastic model by BURRELL [2009]) and important mathematical properties have been fulfilled [GLÄNZEL, 2006]. However, limitations have been found, some of which are shown below.

*This index depends on the scientific field and the number of collaborations.* It is not appropriate, therefore, to compare researchers from different scientific fields, due to different habits of publication, citation, and collaboration. This problem may be corrected, since the maximum $h$ value obtainable in each field strongly correlates with the impact factors of the journals in the field, a reference $h$-index can be estimated in each scientific field [IMPERIAL & RODRÍGUEZ-NAVARRO, 2007]. The $b$-index [BORNMANN ET AL., 2007] is an alternative that indicates the number of articles in the 10% most cited publications in a field, considering ISI-ESI percentiles for example. Multiple authorships and self-citations have been investigated by SCHREIBER [2008a, 2008b]. Concerning the number of collaborators, the $h_1$-index [BATISTA ET AL., 2006] obtained dividing $h$ by the average number of authors of these $h$ articles, can be used.

*This index correlates with the number of publications*. The index tends to favor, therefore, those with more extensive scientific careers and is less effective among those with a low number of publications [CRONIN & MEHO, 2006; SAAD, 2006; VAN-RANN, 2006]. To differentiate between active and inactive researchers and compare scientists at different stages of their careers, the growth rate $h'(t)$ has been proposed, being $t$ the number of years since the publication of the first article [LIANG, 2006; BURRELL, 2007; ROUSSEAU & YE, 2008]. As an alternative, the $h$-index can be calculated for a certain period of time, instead of along the professional life of a researcher.

*All citations of the most cited articles are not considered in this index.* These most cited works contribute to the $h$-index, but their value is not affected by the number of times these articles are cited, since the tails of the citation distribution are not considered. These tails correspond to those publications that move away from the average impact, either because they have been highly cited (upper tail), or less cited (lower tail). Based on the definition of the Hirsch core, several authors have proposed new indicators. The $g$-index [EGGHE, 2006] considers all citations of the $g$ most cited articles, and represents an average citation of these $g$ articles. Once the articles have been sorted in decreasing order of citations, $g$ is the largest value, such that the first $g$ articles have at least $g^2$ citations. As a matter of fact, the $h$-index and the $g$-index are special cases of a family of Hirsch index variants [SCHREIBER, 2010]. Similarly, the $A$-index (average citation) and the $AR$-index (considering the age of the articles) [JIN ET AL., 2007] have the particularity of taking into account the citations of the Hirsch core. However, as stated above, a heavy upper tail may correspond to the work of many authors included in large research lines that generate many citations.

*This index penalizes selective researchers*, that is, those producing a moderate number of high impact articles as opposed to large producers of moderate impact articles. Although this index has proven to be useful in identifying relevant researchers in a field, empirical evidence has shown it does not discriminate among researchers situated at intermediate levels and penalizes selective producers versus large producers [COSTAS & BORDONS, 2007]. Cases with similar values of $h$, where citation curves are intersected, are especially questioned due to some researchers presenting higher levels of citations at the beginning of the curve and lower levels at the end. Additionally, this index is not *consistent* [WALTMAN & VAN-ECK, 2009]. That is, the effect of incorporating a new paper with a given number of citations may be different between researchers, increasing the value of $h$ in some cases and maintaining its value in others.

Finally, some variants have been published in order to improve the *accuracy* of the $h$-index: the tapered $h$-index [ANDERSON ET AL., 2008], the $R_m$-index [PANARETOS & MALESIOS, 2009], the $w$-index [WOHLIN, 2009], and the $e$-index [ZHANG, 2009]. The $h^2$ lower, $h^2$ center, and $h^2$ upper as well as the sRM value [BORNMANN ET AL., 2010], represent new approaches providing additional information that increase the accuracy of the $h$-index.

In this work, a complement to the $h$-index that increases the consistency of the indicator and favors selective authors against large producers is presented. This approach also increases the accuracy of the $h$-index giving information about the shape of the citation distribution. The main difference with respect to the variants which have been proposed previously ($g$-index, $A$-index, $AR$-index, $h^2$ upper, tapered $h$-index, $R_m$-index, $w$-index, and $e$-index) is that all of them are a function of all citations included in a core of most cited papers. By contrast, we establish an upper limit to the maximum number of citations considered for each publication in order to reduce the effect that isolated successes and/or large collaborations may have on the final result (as was pointed out by Hirsh). This upper limit can be modified without further changing the radius of the central index.

## 2. Central indexes

Given the published articles of an author in decreasing order of citations, let $c_i$ be the number of citations received by the publication $i$ $(c_1 \geq c_2 \geq \cdots \geq c_{N_p})$, and let $N_c^j = \sum_{i=1}^{j} c_i$ be the aggregated number of citations of the $j$ most highly cited papers.

The citation distribution is obtained plotting the number of citations versus the position of the articles. Connecting these points, the citation curve is obtained. The $h$-index is the largest integer number that satisfies $c_h \geq h$, that is,

$$h = \max\{i \in Z : c_i \geq i\}.$$

Graphically, the integer part of the intersection point between citation curve and the first quadrant bisector is $h$. This is indicated in Figure 1.

$H = h^2$ is a lower bound for the number of citations of those papers in the Hirsch core. The upper tail $U$ is the excess citations received by the Hirsch core over the lower bound. The lower tail $L$ is the number of citations received by those papers outside the Hirsch core. The following relationships are satisfied:

$$N_c = H + U + L,$$
$$U = N_c^h - H,$$
$$L = N_c - N_c^h.$$

The relative weight of the citation distribution tails is given by $N_c/H$. According to HIRSCH [2005] estimations, if $N_c/H < 3$ the tails of the distribution are light, while if $N_c/H > 5$ the tails are heavy. The $h$-index penalizes those researchers who present heavy tails, especially those with a great tail ratio $U/L$.

An example with the citation curves of two researchers is shown in Figure 1. The first researcher presents higher citation levels at the beginning and lower levels at the end of the curve. Therefore, two different profiles of researchers are appreciated, one more selective and another more massive in the production of papers. However, both scientists have the same $h$-index. A researcher may present less $h$-index than another, although it does not necessarily indicate the former presents a less successful career than the latter. The problem of discriminating between two distributions with similar $h$-index but significantly different distribution tail ratios is presented in Figure 1. As can be appreciated, the higher the rate between tails is presented, the better average citations per article is obtained.

In the above cases, it seems reasonable to measure part of $U$ and $L$ in order to complement the $h$-index with the area around $H$. Thus, the discrimination capacity is increased. This idea allows us to introduce the following index.

*Central area index*

Let $E$ ($F$) be *the upper (lower) area* next to $H$, that is, the part of the upper tail $U$ (lower tail $L$) in the citation distribution closest to $H$. The lower area corresponds to those articles that will likely contribute to increasing the value of $h$ in the future, since they are closer to the Hirsch core. The upper area includes those citations that will form part of $H$ at the time the $h$-index increases its value. Therefore, it seems reasonable to include this area and increase, in this way, the discrimination capacity of the index.

The central area index of radius $j$ is defined as the citations of the $h+j$ most cited papers limited to the number of citations of paper $h-j$. That is, the citations of those papers in the Hirsch core, restricted by the citations of paper $h-j$, jointly with the citations of papers from $h+1$ to $h+j$. The geometrical representation is showed in Figure 2.

The arithmetic definition of the index of radius $j$ is the following:

$$A_j = (h-j)\cdot c_{h-j} + \sum_{i=h-j+1}^{h+j} c_i,\ j = 1,\ldots,h-1.$$

Note that $A_{h-1} = N_c^{2h-1}$ includes the total upper tail $U$. Although the radius could be defined for $j \geq h$, in this case, it would only be adding part of the lower tail.

Figure 3 shows the central area index of radius $j = \lfloor h/2 \rfloor$ (integer part) for two citation distributions. In the case of distribution $D_i$ the central area index is $A_j^i = H + E_i + F_i,\ i = 1,2.$

As mentioned above, authors whose citation distributions have heavy tails are penalized by the $h$-index. However, the central area index of these authors grows faster than those with less heavy tails, increasing its capacity of discrimination.

Selective researchers are also penalized by the $h$-index. However, the central area index solves this problem. For example, suppose a researcher has 10 publications, the least of which have 20 citations. Then $h = 10$, which represents only 100 citations. However, $A_1 \geq 200,$ that is, twice the number of represented citations.

*Central interval index*

The central interval index of radius $j$ is defined as the aggregated citations of the articles from $h$-$j$ to $h$+$j$:

$$I_j = \sum_{i=h-j}^{h+j} c_i,\ j = 1,\ldots,h-1.$$

That is, the citations at the interval $[h-j,\ h+j]$. Note that $I_{h-1} = A_{h-1} = N_c^{2h-1}$. The geometrical representation is shown in Figure 2.

The central interval index of radius $j = \lfloor h/2 \rfloor$ (integer part) for two citation distributions is shown in Figure 3. In the case of distribution $D_i$ the central interval index is $I_j^i = G_i + F_i,\ i = 1,2.$

*Comparison among central indexes.*

Both indexes have the same area in the lower tail. However, significant differences between both indexes are appreciated in the upper tail. Thus while central interval indexes add citations of articles to the left of $h$, the central area indexes add zones of variable size in the upper tail.

Lets now look at differences between both indexes for two authors with the same $h$, one more selective than the other. Increasing the radius in one unit and reducing the comparison to the upper tail, where differences exist, the central interval index adds to the selective author the height of the rectangle $R$, as shown Figure 3. However, the central area index also adds to the selective author the area $R$. For this reason, the central area index is more beneficial for selective authors. As an author is more selective, the height of $R$ increases and the area of $R$ also increases in a greater proportion.

In the following section, an empirical application determines an optimal radius of the central indexes, obtaining the value of $j$ that best describes the central shape of the citation distribution.

## 3. Empirical application

The behavior of central indexes for researchers who have received the Price Medal is analyzed in this section. Data about these scientists was obtained from the ISI Web of Science database in February 2010. To estimate the predictive capacity of indexes for five and ten years ahead, and their comparison with the $h$-index, the cited articles and the number of citations obtained in 1999, 2004, and 2009, have been considered. In order not to distort further analysis, especially regression analysis, only the 15 existing and currently productive scientists were considered.

The objective consists in obtaining $A_j,\ I_j,\ j = 1,\ldots,h-1,$ at instant $t$ and estimating the value of $j$ (optimal radius) most correlated with $A_k,\ I_k,\ k = j,\ldots,h-1,$ at instant $t+1,\ t = 1,2,$ and $t+2,\ t = 1$, that is, the future indexes.

Table 1 shows, for each author, the year of the first article published in the database, the total number of cited articles and the total citations in 1999, 2004, and 2009. This table also shows the evolution of the $h$-index.

Figure 4 shows the citation curves of four researchers. Three curves are shown, the closest to the origin corresponds to 1999, followed by 2004 and the farthest to 2009. The value 100 has been taken as maximum only for clarity. This plot allows us to observe the evolution of the $h$-index, and also to distinguish between selective and large producer researchers. As an example, McCain and Small show a more selective behavior than Egghe and Garfield, respectively.

Production-impact scatter plots are presented in Figure 5. As shown, linear correlation between the number of articles and the number of citations exists. Authors located above the regression line show a more selective behavior than those below this line. Thus, the more selective authors of the sample are Small and Garfield, respectively.

Table 2 shows the central indexes of the years considered. These indexes can be obtained up to a radius of 24 for some authors, but as the radius increases the number of data in each column of the table is reduced. Data has been shown until radius 10 to ensure that later the correlation coefficient is calculated with more than half of the sample data (at least 9 out of 15). This table is useful in estimating future success. Lets see some examples using radius 7 as reference, approximately half the average $h$-index for the first period. It will be shown later that this indicator provides good estimations for five year predictions. As can be seen in the case of Leydesdorff, the area index varies from $A_1 = 97$ to $A_7 = 171$ in 1999, a significant increment that reveals the evolution of the $h$-index in following periods (2004 and 2009). Indeed, this author has an $h$-index of 9, 13, and 21, respectively. Something similar can also be seen in the case of McCain, among others. These examples suggest the area indexes obtained in a period, predict the increase in the following period.

Lets now see a comparison between two authors with the same $h$. McCain and Vlachy have $h_{1999} = 11$. $A_6$ in the case of McCain (because $A_7$ is not defined) is greater than the case of Vlachy, which estimates a higher future $h$-index; which holds true (15 vs. 11) in Table 1. Something similar happens with Ingwersen and Vinkler, for which $h_{1999} = 7$. $A_7$ in the case of Ingwersen is greater than in the case of Vinkler, which estimates a higher future $h$-index; which also holds true (12 vs. 10) in Table 1. The same conclusion can also be observed for a period of ten years. Although area indexes have been taken as reference, something similar occurs in the case of interval indexes.

Now, as an example, lets consider a case where the discrimination capacity of the central indexes compared to the $h$-index is appreciated. Braun has $h_{1999}$ greater than Small. However, the central indexes from a certain radius are higher for the later author. Attending to these indicators, the second author seems more selective, which is true according to total citations and the production-impact scatter plot.

Since the central index is an aggregation of citations, its representation with respect to the radius is an increasing function, as can be seen in Figure 6. The first plot shows that McCain's area indexes are higher than those of Egghe, indicating the first author is more selective than the second. Something similar can be seen with the interval index in the second plot.

Table 3 shows the linear correlation coefficients among indexes for 5 and 10 years. Matrices of order 10 to ensure the correlation coefficient is calculated with more than half of the researchers (at least 9 out of 15) are shown. As can be seen, the area indexes for 1999 are strongly correlated with those for 2004, so they look like good estimators for 5 years. In all cases, correlations are higher than 0.94. The strongest correlations are located close to the main diagonal of the matrix. From the fifth element, all coefficients on the diagonal are greater than the correlation between $h$-indexes $\mathrm{corr}(h_{1999}, h_{2004}) = 0.977$. As can be seen, all elements in column 7 are also higher than this. Therefore, $A_7$ seems a good estimator for five years and the radius is about half the average $h$-index of the sample.

Area indexes for 1999 also show high correlations with 2009, although slightly lower than those mentioned in the previous paragraph, making them also good estimators for 10 years. All of the coefficients are greater than the correlation between the $h$-indexes $\mathrm{corr}(h_{1999}, h_{2009}) = 0.812$. As can be seen, all elements in column 7 are higher than 0.9.

Finally, the area indexes for 2004 also present correlations with the year 2009. Most of the elements (including all of them in column 7) are higher than the correlation between $h$-indexes $\mathrm{corr}(h_{2004}, h_{2009}) = 0.889$.

With respect to interval indexes, something very similar occurs. Correlations are also high in all cases. In order to better appreciate what indicators provide the best correlations, the differences between correlations for central indexes are also shown in Table 3. As can be seen, most of the elements of these matrices are positive, which means the correlations for the area index are greater than for the interval index (only 10 out of 165 items are negative).

## 4. Conclusions

The h-index is a bibliometric indicator that attempts to measure the success of a researcher with just a part of the total amount of publications and citations. Due to not considering all production and impact, this index corrects biases of mass collaborations and punctual successes, which may not be significant in the researcher's career as a whole. However, different citation distributions, like those of a selective researcher and a large producer, may cause similar h-indexes, and in these cases, it is not possible to distinguish between these researchers using the h-index exclusively.

In this paper two complements to the h-index, the area and the interval indexes, have been proposed with the aim of increasing the capacity of discrimination among researchers with similar h, and improving the prediction of future successes. These indicators consider some areas that are larger for selective authors than for large producers. Thus, a problem described in the literature about the h-index, which penalizes selective researchers compared to large producers, is corrected.

Both central indexes are good estimators and correlations are generally higher for the area index than for the interval index. Moreover, a radius that well describes the shape of the citation distribution has been estimated empirically. This radius is about half the average h-index of researchers being evaluated.

Finally, we would like to point out that the area index is not considered a substitute, but a complement to the h-index, especially in an evaluation process where doubts among researchers might exist.

## Materials and Methods

The publications and citations for those scientists listed in Table 1 were obtained from ISI Web of Science database in February, 2010.

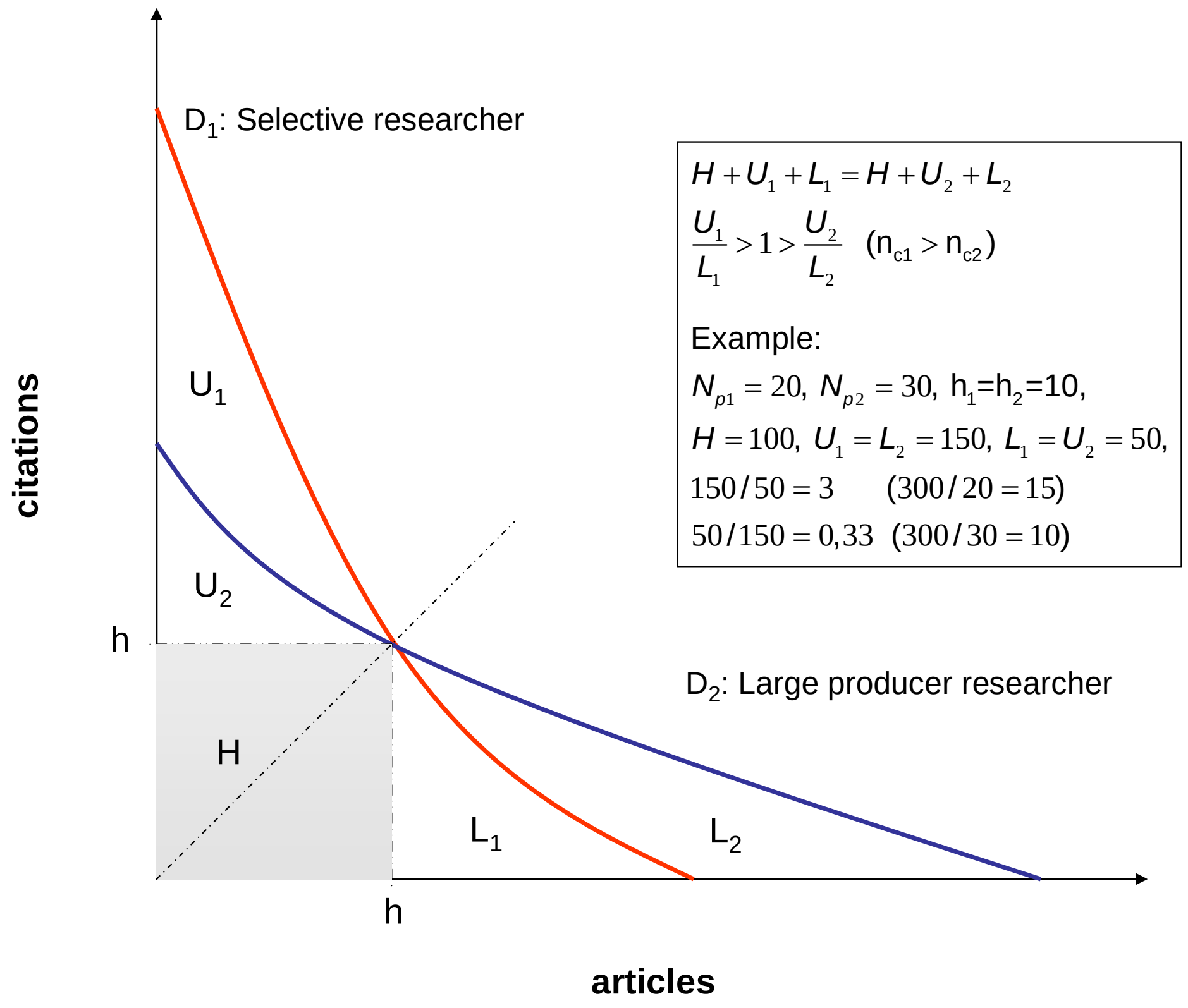

Figure 1. Two citation curves with the same h-index but different average citations per article.

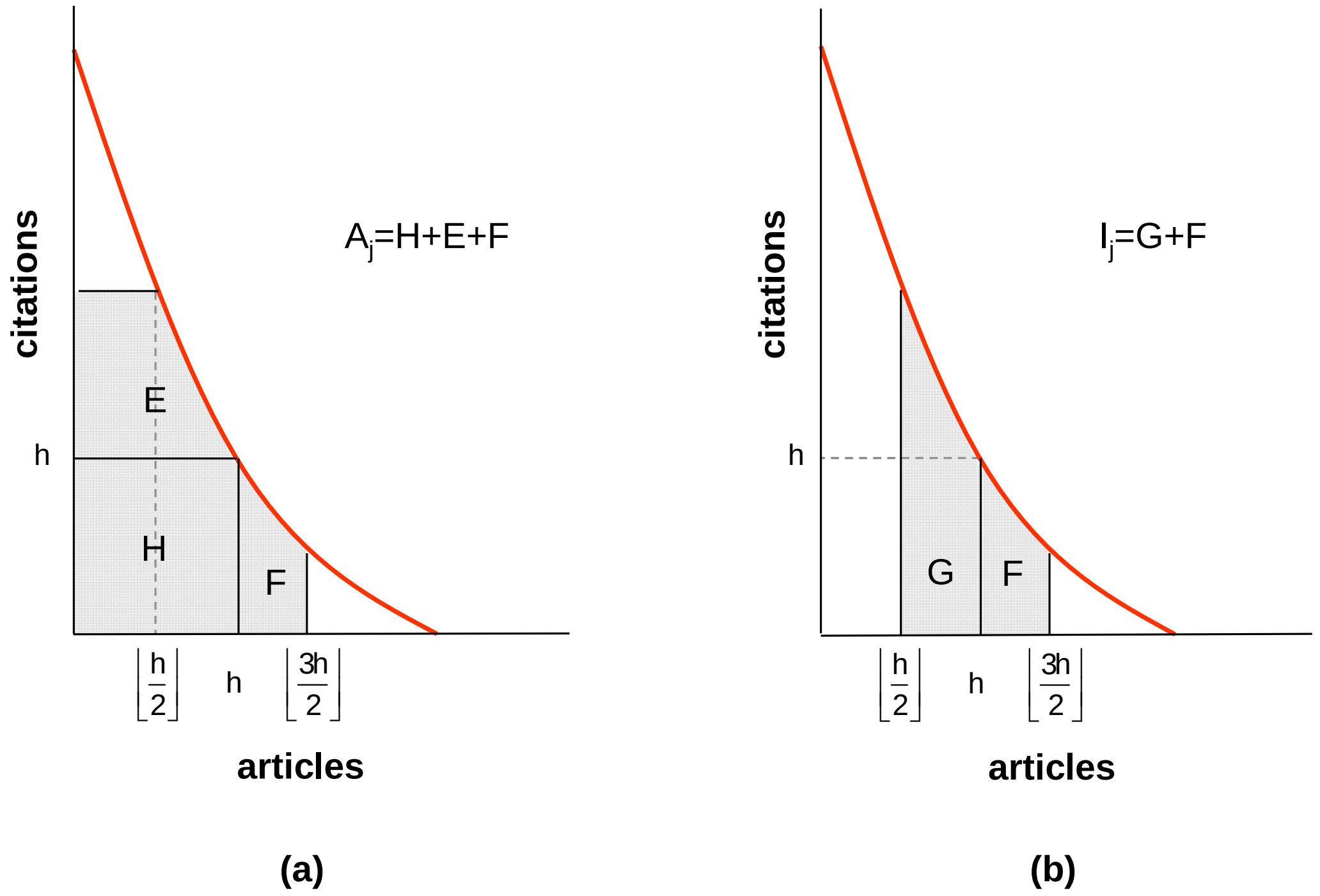

Figure 2. Central area index (a) and central interval index (b) of radius $j = \left\lfloor \frac{h}{2} \right\rfloor$.

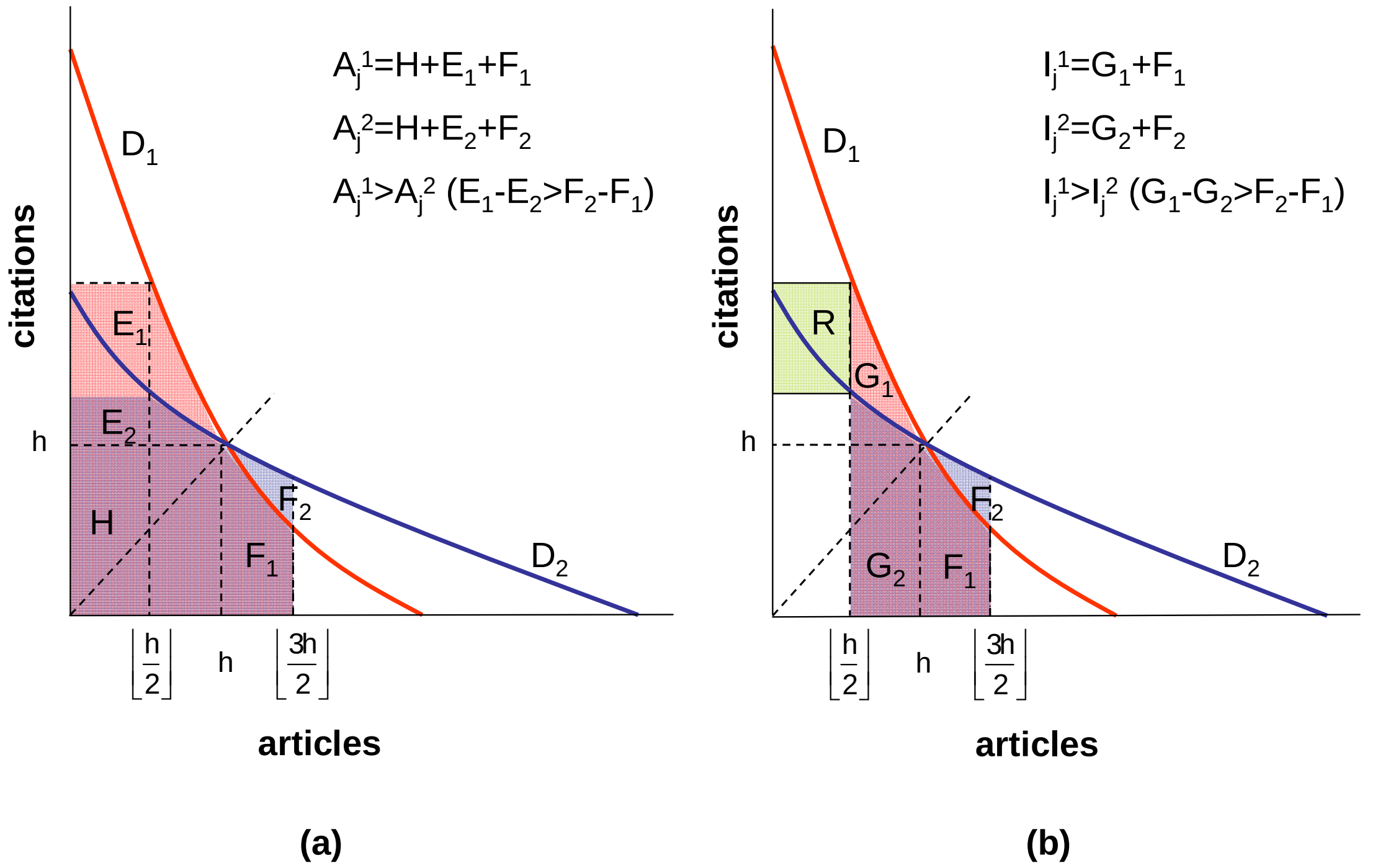

Figure 3. Comparison between the central area index (a) and the central interval index (b) of radius $j = \left\lfloor \frac{h}{2} \right\rfloor$, for two citation curves with the same h-index.

Table 1. The production and impact of the researchers.

| Author | Year of the first article | Cited articles | | | Citations | | | h-index | | | $H=h^2$ | | |
|---|---|---|---|---|---|---|---|---|---|---|---|---|---|
| | | 1999 | 2004 | 2009 | 1999 | 2004 | 2009 | 1999 | 2004 | 2009 | 1999 | 2004 | 2009 |
| Braun, T | 1958 | 135 | 152 | 170 | 1966 | 2498 | 3116 | 24 | 27 | 30 | 576 | 729 | 900 |
| Egghe, L | 1978 | 47 | 78 | 122 | 299 | 571 | 1277 | 10 | 12 | 18 | 100 | 144 | 324 |
| Garfield, E | 1954 | 163 | 174 | 180 | 3687 | 4298 | 5294 | 25 | 26 | 29 | 625 | 676 | 841 |
| Glänzel, W | 1983 | 52 | 74 | 112 | 616 | 991 | 2228 | 14 | 18 | 28 | 196 | 324 | 784 |
| Ingwersen, P | 1982 | 18 | 27 | 35 | 239 | 686 | 1160 | 7 | 12 | 16 | 49 | 144 | 256 |
| Leydesdorff, L | 1981 | 38 | 54 | 107 | 235 | 477 | 1541 | 9 | 13 | 21 | 81 | 169 | 441 |
| McCain, KW | 1983 | 25 | 32 | 40 | 328 | 761 | 1261 | 11 | 15 | 17 | 121 | 225 | 289 |
| Moed, HF | 1985 | 31 | 50 | 64 | 386 | 804 | 1608 | 12 | 16 | 22 | 144 | 256 | 484 |
| Rousseau, R | 1986 | 40 | 76 | 122 | 165 | 494 | 1339 | 6 | 11 | 20 | 36 | 121 | 400 |
| Schubert, A | 1981 | 75 | 104 | 121 | 726 | 1126 | 1904 | 14 | 18 | 24 | 196 | 324 | 576 |
| Small, H | 1961 | 59 | 64 | 69 | 2947 | 3543 | 4296 | 21 | 24 | 25 | 441 | 576 | 625 |
| Van-Raan, AFJ | 1976 | 47 | 64 | 78 | 488 | 909 | 1750 | 13 | 17 | 24 | 169 | 289 | 576 |
| Vinkler, P | 1986 | 20 | 26 | 31 | 149 | 266 | 411 | 7 | 10 | 13 | 49 | 100 | 169 |
| Vlachy, J | 1963 | 41 | 42 | 43 | 361 | 374 | 382 | 11 | 11 | 11 | 121 | 121 | 121 |
| Zitt, M | 1991 | 6 | 12 | 23 | 17 | 78 | 267 | 3 | 6 | 10 | 9 | 36 | 100 |
| Average | 1977 | 53,1 | 68,6 | 87,8 | 840,6 | 1191,7 | 1855,6 | 12,5 | 15,7 | 20,5 | 194,2 | 282,3 | 459,1 |

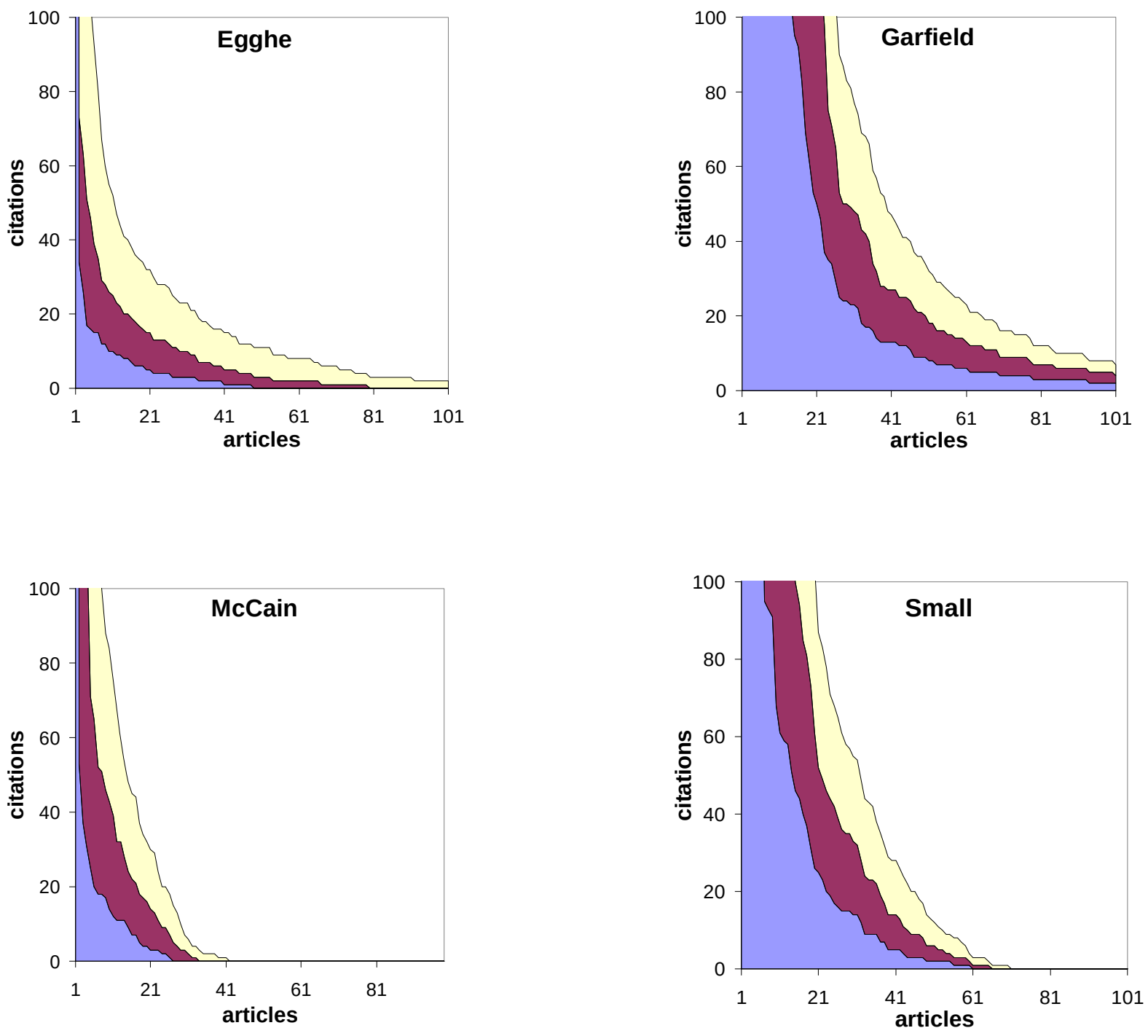

Figure 4. Citation curves for some researchers in 1999, 2004, and 2009.

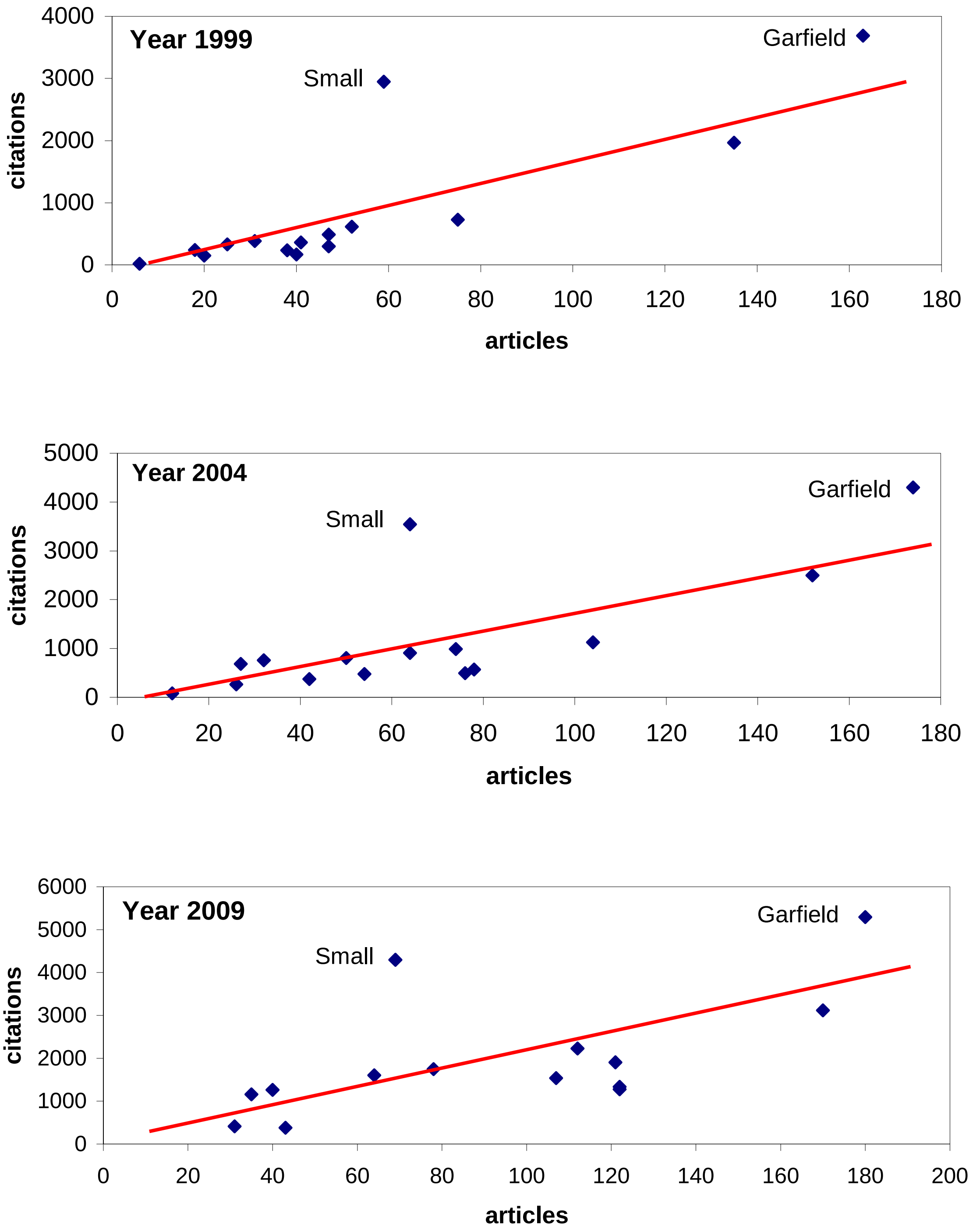

Figure 5. Production-impact scatter plot.

Table 2. Central indexes up to radius 10.

| Year | Author | $A_1$ | $A_2$ | $A_3$ | $A_4$ | $A_5$ | $A_6$ | $A_7$ | $A_8$ | $A_9$ | $A_{10}$ | $I_1$ | $I_2$ | $I_3$ | $I_4$ | $I_5$ | $I_6$ | $I_7$ | $I_8$ | $I_9$ | $I_{10}$ |
|---|---|---|---|---|---|---|---|---|---|---|---|---|---|---|---|---|---|---|---|---|---|
| 1999 | Braun, T | 597 | 618 | 660 | 701 | 722 | 797 | 817 | 869 | 903 | 950 | 69 | 114 | 160 | 207 | 254 | 304 | 353 | 404 | 455 | 508 |
| | Egghe, L | 109 | 134 | 142 | 168 | 175 | 185 | 194 | 218 | 231 | - | 29 | 50 | 70 | 93 | 115 | 137 | 160 | 192 | 231 | - |
| | Garfield, E | 870 | 917 | 985 | 1197 | 1300 | 1379 | 1541 | 1694 | 1935 | 2086 | 88 | 147 | 208 | 277 | 350 | 425 | 504 | 590 | 690 | 798 |
| | Glänzel, W | 237 | 275 | 298 | 330 | 341 | 352 | 369 | 384 | 398 | 411 | 45 | 77 | 108 | 141 | 173 | 205 | 237 | 269 | 302 | 336 |
| | Ingwersen, P | 70 | 106 | 155 | 166 | 211 | 230 | - | - | - | - | 25 | 46 | 77 | 110 | 163 | 230 | - | - | - | - |
| | Leydesdorff, L | 97 | 112 | 119 | 136 | 155 | 161 | 171 | 179 | - | - | 27 | 46 | 64 | 84 | 107 | 129 | 153 | 179 | - | - |
| | McCain, KW | 142 | 171 | 204 | 218 | 225 | 240 | 264 | 286 | 301 | 320 | 34 | 59 | 85 | 110 | 135 | 160 | 189 | 224 | 264 | 320 |
| | Moed, HF | 167 | 186 | 221 | 228 | 249 | 267 | 283 | 293 | 297 | 315 | 37 | 60 | 85 | 109 | 135 | 162 | 191 | 221 | 249 | 284 |
| | Rousseau, R | 53 | 63 | 69 | 80 | 88 | - | - | - | - | - | 21 | 36 | 51 | 68 | 88 | - | - | - | - | - |
| | Schubert, A | 223 | 248 | 282 | 324 | 336 | 355 | 372 | 388 | 403 | 428 | 43 | 72 | 102 | 135 | 168 | 201 | 234 | 268 | 303 | 341 |
| | Small, H | 543 | 581 | 688 | 806 | 869 | 944 | 987 | 1066 | 1164 | 1187 | 68 | 113 | 161 | 214 | 269 | 328 | 389 | 454 | 526 | 597 |
| | Van-Raan, AFJ | 182 | 194 | 226 | 255 | 265 | 274 | 283 | 302 | 315 | 335 | 39 | 64 | 91 | 119 | 146 | 172 | 198 | 226 | 255 | 287 |
| | Vinkler, P | 87 | 96 | 119 | 124 | 132 | 140 | - | - | - | - | 27 | 44 | 65 | 86 | 110 | 140 | - | - | - | - |
| | Vlachy, J | 153 | 162 | 187 | 196 | 222 | 230 | 246 | 262 | 275 | 286 | 36 | 58 | 82 | 106 | 132 | 158 | 186 | 216 | 249 | 286 |
| | Zitt, M | 13 | 15 | - | - | - | - | - | - | - | - | 9 | 15 | - | - | - | - | - | - | - | - |
| 2004 | Braun, T | 755 | 805 | 853 | 876 | 921 | 964 | 986 | 1046 | 1085 | 1123 | 80 | 133 | 186 | 238 | 291 | 344 | 397 | 452 | 507 | 563 |
| | Egghe, L | 179 | 201 | 222 | 234 | 252 | 280 | 310 | 344 | 365 | 380 | 39 | 66 | 94 | 122 | 150 | 180 | 214 | 254 | 297 | 343 |
| | Garfield, E | 954 | 1004 | 1099 | 1586 | 1716 | 2001 | 2064 | 2105 | 2259 | 2389 | 90 | 153 | 219 | 305 | 396 | 500 | 606 | 711 | 819 | 934 |
| | Glänzel, W | 342 | 376 | 393 | 423 | 464 | 491 | 505 | 549 | 571 | 583 | 54 | 91 | 127 | 163 | 200 | 238 | 275 | 315 | 355 | 394 |
| | Ingwersen, P | 213 | 224 | 332 | 356 | 369 | 417 | 473 | 575 | 611 | 636 | 43 | 71 | 108 | 146 | 183 | 227 | 281 | 359 | 447 | 543 |
| | Leydesdorff, L | 178 | 209 | 228 | 254 | 262 | 284 | 298 | 306 | 318 | 328 | 35 | 59 | 84 | 110 | 136 | 164 | 193 | 222 | 252 | 282 |
| | McCain, KW | 239 | 278 | 339 | 351 | 422 | 450 | 458 | 493 | 506 | 567 | 44 | 74 | 108 | 141 | 179 | 218 | 255 | 295 | 336 | 387 |
| | Moed, HF | 318 | 348 | 375 | 388 | 434 | 447 | 469 | 513 | 585 | 605 | 52 | 88 | 123 | 157 | 194 | 231 | 269 | 310 | 357 | 405 |
| | Rousseau, R | 142 | 152 | 162 | 179 | 195 | 209 | 222 | 231 | 248 | 258 | 34 | 56 | 78 | 101 | 125 | 149 | 174 | 199 | 228 | 258 |
| | Schubert, A | 341 | 357 | 373 | 401 | 428 | 465 | 489 | 501 | 522 | 550 | 53 | 87 | 121 | 154 | 188 | 223 | 259 | 294 | 330 | 368 |
| | Small, H | 623 | 666 | 686 | 726 | 897 | 1041 | 1091 | 1122 | 1211 | 1281 | 73 | 120 | 166 | 213 | 267 | 327 | 387 | 447 | 511 | 579 |
| | Van-Raan, AFJ | 306 | 337 | 353 | 394 | 444 | 469 | 482 | 504 | 525 | 545 | 51 | 85 | 119 | 154 | 191 | 229 | 266 | 304 | 343 | 383 |
| | Vinkler, P | 148 | 163 | 169 | 193 | 198 | 219 | 229 | 247 | 254 | - | 36 | 58 | 79 | 103 | 126 | 153 | 181 | 216 | 254 | - |
| | Vlachy, J | 153 | 171 | 188 | 197 | 223 | 236 | 252 | 268 | 279 | 294 | 36 | 59 | 83 | 107 | 133 | 160 | 189 | 220 | 253 | 294 |
| | Zitt, M | 52 | 57 | 64 | 73 | 76 | - | - | - | - | - | 20 | 33 | 46 | 61 | 76 | - | - | - | - | - |
| 2009 | Braun, T | 958 | 986 | 1014 | 1067 | 1144 | 1218 | 1266 | 1312 | 1356 | 1379 | 90 | 149 | 208 | 267 | 328 | 390 | 452 | 514 | 576 | 638 |
| | Egghe, L | 341 | 358 | 390 | 419 | 447 | 474 | 511 | 555 | 586 | 623 | 53 | 88 | 124 | 159 | 195 | 232 | 271 | 312 | 354 | 399 |
| | Garfield, E | 985 | 1120 | 1146 | 1222 | 1368 | 1439 | 2124 | 2590 | 2674 | 2733 | 94 | 158 | 221 | 286 | 356 | 427 | 528 | 650 | 774 | 897 |
| | Glänzel, W | 812 | 839 | 866 | 916 | 964 | 988 | 1033 | 1097 | 1120 | 1196 | 84 | 139 | 194 | 249 | 304 | 358 | 413 | 470 | 526 | 584 |
| | Ingwersen, P | 301 | 342 | 394 | 417 | 504 | 523 | 577 | 586 | 649 | 733 | 49 | 82 | 118 | 153 | 194 | 235 | 281 | 327 | 379 | 443 |
| | Leydesdorff, L | 461 | 557 | 577 | 611 | 627 | 703 | 732 | 760 | 785 | 798 | 62 | 107 | 152 | 195 | 237 | 283 | 329 | 376 | 422 | 468 |
| | McCain, KW | 404 | 420 | 450 | 479 | 528 | 616 | 637 | 693 | 711 | 769 | 59 | 98 | 138 | 179 | 220 | 266 | 313 | 365 | 417 | 475 |
| | Moed, HF | 548 | 609 | 629 | 703 | 791 | 827 | 876 | 892 | 919 | 969 | 68 | 115 | 161 | 210 | 263 | 317 | 372 | 424 | 475 | 529 |
| | Rousseau, R | 438 | 456 | 491 | 508 | 540 | 570 | 586 | 614 | 630 | 666 | 60 | 99 | 139 | 178 | 218 | 258 | 298 | 339 | 380 | 423 |
| | Schubert, A | 623 | 647 | 712 | 734 | 755 | 793 | 813 | 833 | 868 | 901 | 73 | 122 | 172 | 221 | 269 | 317 | 365 | 413 | 462 | 511 |
| | Small, H | 675 | 721 | 853 | 917 | 959 | 1132 | 1170 | 1309 | 1456 | 1502 | 77 | 127 | 181 | 237 | 294 | 358 | 422 | 493 | 571 | 648 |
| | Van-Raan, AFJ | 648 | 671 | 715 | 758 | 856 | 896 | 934 | 955 | 990 | 1022 | 76 | 125 | 175 | 226 | 280 | 335 | 390 | 445 | 500 | 554 |
| | Vinkler, P | 219 | 241 | 259 | 276 | 292 | 300 | 324 | 344 | 353 | 361 | 43 | 71 | 97 | 124 | 152 | 180 | 209 | 240 | 272 | 305 |
| | Vlachy, J | 153 | 173 | 190 | 199 | 225 | 238 | 254 | 270 | 281 | 299 | 36 | 61 | 85 | 109 | 135 | 162 | 191 | 222 | 255 | 299 |
| | Zitt, M | 137 | 161 | 169 | 193 | 204 | 221 | 230 | 237 | 261 | - | 33 | 56 | 79 | 103 | 128 | 155 | 182 | 211 | 261 | - |

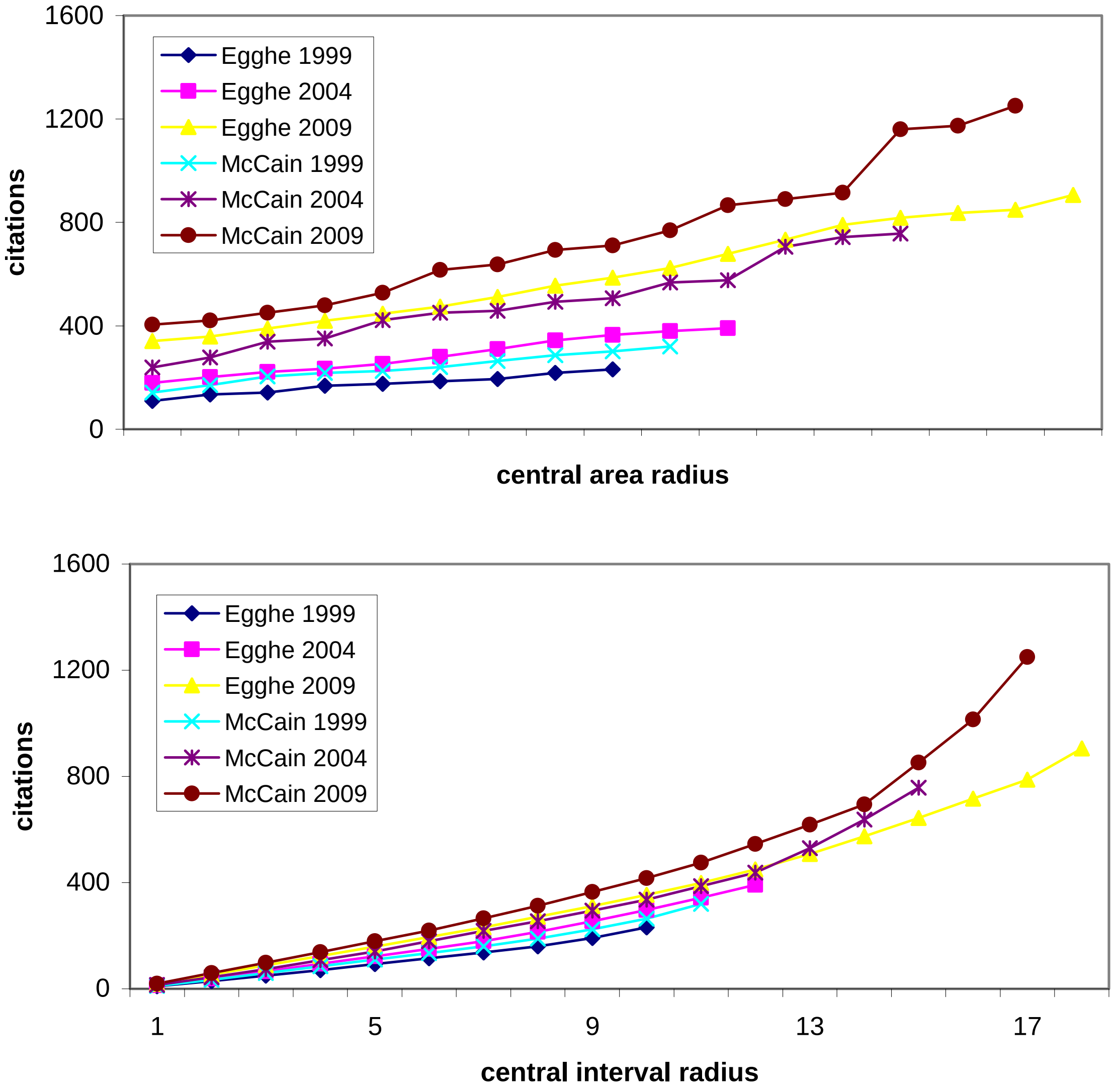

Figure 6. Comparison of central indexes for two authors.

# Table 3. Indexes correlations.

| | | 1999 | | | | | | | | | |
|---|---|---|---|---|---|---|---|---|---|---|---|
| | | $A_1$ | $A_2$ | $A_3$ | $A_4$ | $A_5$ | $A_6$ | $A_7$ | $A_8$ | $A_9$ | $A_{10}$ |
| | $A_1$ | 0,986 | | | | | | | | | |
| | $A_2$ | 0,985 | 0,987 | | | | | | | | |
| | $A_3$ | 0,977 | 0,980 | 0,978 | | | | | | | |
| | $A_4$ | 0,967 | 0,970 | 0,962 | 0,967 | | | | | | |
| 2004 | $A_5$ | 0,973 | 0,977 | 0,973 | 0,979 | 0,981 | | | | | |
| | $A_6$ | 0,964 | 0,969 | 0,967 | 0,978 | 0,981 | 0,977 | | | | |
| | $A_7$ | 0,961 | 0,966 | 0,966 | 0,977 | 0,981 | 0,977 | 0,986 | | | |
| | $A_8$ | 0,956 | 0,962 | 0,963 | 0,973 | 0,977 | 0,974 | 0,986 | 0,987 | | |
| | $A_9$ | 0,952 | 0,959 | 0,960 | 0,971 | 0,976 | 0,972 | 0,985 | 0,986 | 0,988 | |
| | $A_{10}$ | 0,948 | 0,955 | 0,957 | 0,968 | 0,974 | 0,970 | 0,984 | 0,986 | 0,988 | 0,990 |
| | Average | 0,967 | 0,969 | 0,966 | 0,973 | 0,978 | 0,974 | 0,985 | 0,986 | 0,988 | 0,990 |

| | | 1999 | | | | | | | | | |
|---|---|---|---|---|---|---|---|---|---|---|---|
| | | $I_1$ | $I_2$ | $I_3$ | $I_4$ | $I_5$ | $I_6$ | $I_7$ | $I_8$ | $I_9$ | $I_{10}$ |
| | $I_1$ | 0,977 | | | | | | | | | |
| | $I_2$ | 0,976 | 0,979 | | | | | | | | |
| | $I_3$ | 0,970 | 0,975 | 0,975 | | | | | | | |
| | $I_4$ | 0,965 | 0,972 | 0,974 | 0,975 | | | | | | |
| 2004 | $I_5$ | 0,962 | 0,969 | 0,971 | 0,974 | 0,973 | | | | | |
| | $I_6$ | 0,947 | 0,956 | 0,966 | 0,970 | 0,973 | 0,962 | | | | |
| | $I_7$ | 0,934 | 0,945 | 0,957 | 0,964 | 0,971 | 0,964 | 0,967 | | | |
| | $I_8$ | 0,911 | 0,924 | 0,941 | 0,951 | 0,965 | 0,965 | 0,964 | 0,967 | | |
| | $I_9$ | 0,884 | 0,899 | 0,920 | 0,933 | 0,954 | 0,961 | 0,961 | 0,964 | 0,963 | |
| | $I_{10}$ | 0,850 | 0,866 | 0,893 | 0,910 | 0,937 | 0,949 | 0,957 | 0,961 | 0,962 | 0,966 |
| | Average | 0,938 | 0,943 | 0,950 | 0,954 | 0,962 | 0,960 | 0,962 | 0,964 | 0,963 | 0,966 |

| | | 1999 subtraction of correlations (area minus interval) | | | | | | | | | |
|---|---|---|---|---|---|---|---|---|---|---|---|
| | | 0,009 | | | | | | | | | |
| | | 0,009 | 0,007 | | | | | | | | |
| | | 0,007 | 0,005 | 0,003 | | | | | | | |
| | | 0,002 | -0,001 | -0,012 | -0,008 | | | | | | |
| 2004 | | 0,011 | 0,008 | 0,002 | 0,005 | 0,007 | | | | | |
| | | 0,017 | 0,013 | 0,002 | 0,007 | 0,008 | 0,016 | | | | |
| | | 0,026 | 0,021 | 0,009 | 0,012 | 0,010 | 0,013 | 0,019 | | | |
| | | 0,045 | 0,039 | 0,022 | 0,022 | 0,012 | 0,009 | 0,021 | 0,020 | | |
| | | 0,068 | 0,060 | 0,040 | 0,037 | 0,022 | 0,012 | 0,024 | 0,022 | 0,024 | |
| | | 0,098 | 0,088 | 0,064 | 0,059 | 0,037 | 0,021 | 0,027 | 0,025 | 0,026 | 0,024 |
| | Sum | 0,292 | 0,240 | 0,130 | 0,135 | 0,096 | 0,070 | 0,090 | 0,067 | 0,050 | 0,024 |

| | | 1999 | | | | | | | | | |
|---|---|---|---|---|---|---|---|---|---|---|---|
| | | $A_1$ | $A_2$ | $A_3$ | $A_4$ | $A_5$ | $A_6$ | $A_7$ | $A_8$ | $A_9$ | $A_{10}$ |
| | $A_1$ | 0,815 | | | | | | | | | |
| | $A_2$ | 0,836 | 0,841 | | | | | | | | |
| | $A_3$ | 0,843 | 0,849 | 0,830 | | | | | | | |
| | $A_4$ | 0,846 | 0,852 | 0,833 | 0,826 | | | | | | |
| 2009 | $A_5$ | 0,853 | 0,859 | 0,842 | 0,835 | 0,825 | | | | | |
| | $A_6$ | 0,864 | 0,871 | 0,860 | 0,853 | 0,843 | 0,840 | | | | |
| | $A_7$ | 0,914 | 0,920 | 0,905 | 0,913 | 0,910 | 0,906 | 0,906 | | | |
| | $A_8$ | 0,922 | 0,928 | 0,913 | 0,926 | 0,926 | 0,922 | 0,926 | 0,926 | | |
| | $A_9$ | 0,928 | 0,934 | 0,923 | 0,936 | 0,937 | 0,933 | 0,937 | 0,937 | 0,943 | |
| | $A_{10}$ | 0,916 | 0,924 | 0,918 | 0,931 | 0,933 | 0,928 | 0,934 | 0,934 | 0,938 | 0,938 |
| | Average | 0,874 | 0,886 | 0,878 | 0,889 | 0,895 | 0,906 | 0,926 | 0,933 | 0,940 | 0,938 |

| | | 1999 | | | | | | | | | |
|---|---|---|---|---|---|---|---|---|---|---|---|
| | | $I_1$ | $I_2$ | $I_3$ | $I_4$ | $I_5$ | $I_6$ | $I_7$ | $I_8$ | $I_9$ | $I_{10}$ |
| | $I_1$ | 0,820 | | | | | | | | | |
| | $I_2$ | 0,820 | 0,826 | | | | | | | | |
| | $I_3$ | 0,818 | 0,825 | 0,770 | | | | | | | |
| | $I_4$ | 0,822 | 0,829 | 0,777 | 0,768 | | | | | | |
| 2009 | $I_5$ | 0,829 | 0,836 | 0,787 | 0,779 | 0,755 | | | | | |
| | $I_6$ | 0,836 | 0,843 | 0,797 | 0,790 | 0,767 | 0,732 | | | | |
| | $I_7$ | 0,857 | 0,864 | 0,827 | 0,822 | 0,803 | 0,771 | 0,800 | | | |
| | $I_8$ | 0,878 | 0,886 | 0,857 | 0,853 | 0,838 | 0,810 | 0,841 | 0,834 | | |
| | $I_9$ | 0,888 | 0,897 | 0,877 | 0,876 | 0,865 | 0,840 | 0,869 | 0,864 | 0,858 | |
| | $I_{10}$ | 0,880 | 0,891 | 0,891 | 0,892 | 0,884 | 0,864 | 0,889 | 0,886 | 0,882 | 0,868 |
| | Average | 0,845 | 0,855 | 0,823 | 0,826 | 0,819 | 0,803 | 0,850 | 0,862 | 0,870 | 0,868 |

| | | 1999 subtraction of correlations (area minus interval) | | | | | | | | | |
|---|---|---|---|---|---|---|---|---|---|---|---|
| | | -0,005 | | | | | | | | | |
| | | 0,016 | 0,014 | | | | | | | | |
| | | 0,025 | 0,024 | 0,059 | | | | | | | |
| | | 0,024 | 0,023 | 0,056 | 0,058 | | | | | | |
| 2009 | | 0,024 | 0,023 | 0,055 | 0,056 | 0,070 | | | | | |
| | | 0,029 | 0,028 | 0,062 | 0,063 | 0,076 | 0,108 | | | | |
| | | 0,057 | 0,056 | 0,078 | 0,091 | 0,107 | 0,135 | 0,106 | | | |
| | | 0,044 | 0,042 | 0,056 | 0,073 | 0,088 | 0,112 | 0,085 | 0,092 | | |
| | | 0,040 | 0,038 | 0,046 | 0,060 | 0,072 | 0,092 | 0,068 | 0,073 | 0,085 | |
| | | 0,036 | 0,032 | 0,027 | 0,040 | 0,048 | 0,064 | 0,044 | 0,048 | 0,056 | 0,071 |
| | Sum | 0,291 | 0,280 | 0,440 | 0,441 | 0,461 | 0,511 | 0,303 | 0,213 | 0,141 | 0,071 |

| | | 2004 | | | | | | | | | |
|---|---|---|---|---|---|---|---|---|---|---|---|
| | | $A_1$ | $A_2$ | $A_3$ | $A_4$ | $A_5$ | $A_6$ | $A_7$ | $A_8$ | $A_9$ | $A_{10}$ |
| | $A_1$ | 0,872 | | | | | | | | | |
| | $A_2$ | 0,889 | 0,893 | | | | | | | | |
| | $A_3$ | 0,896 | 0,899 | 0,883 | | | | | | | |
| | $A_4$ | 0,899 | 0,903 | 0,886 | 0,842 | | | | | | |
| 2009 | $A_5$ | 0,908 | 0,911 | 0,901 | 0,864 | 0,865 | | | | | |
| | $A_6$ | 0,916 | 0,920 | 0,908 | 0,863 | 0,871 | 0,844 | | | | |
| | $A_7$ | 0,940 | 0,941 | 0,941 | 0,953 | 0,953 | 0,943 | 0,940 | | | |
| | $A_8$ | 0,934 | 0,934 | 0,936 | 0,967 | 0,968 | 0,963 | 0,960 | 0,956 | | |
| | $A_9$ | 0,939 | 0,939 | 0,941 | 0,969 | 0,972 | 0,968 | 0,966 | 0,962 | 0,963 | |
| | $A_{10}$ | 0,928 | 0,928 | 0,933 | 0,964 | 0,967 | 0,966 | 0,965 | 0,962 | 0,963 | 0,962 |
| | Average | 0,912 | 0,919 | 0,916 | 0,917 | 0,933 | 0,937 | 0,958 | 0,960 | 0,963 | 0,962 |

| | | 2004 | | | | | | | | | |
|---|---|---|---|---|---|---|---|---|---|---|---|
| | | $I_1$ | $I_2$ | $I_3$ | $I_4$ | $I_5$ | $I_6$ | $I_7$ | $I_8$ | $I_9$ | $I_{10}$ |
| | $I_1$ | 0,875 | | | | | | | | | |
| | $I_2$ | 0,873 | 0,879 | | | | | | | | |
| | $I_3$ | 0,872 | 0,878 | 0,871 | | | | | | | |
| | $I_4$ | 0,877 | 0,883 | 0,875 | 0,857 | | | | | | |
| 2009 | $I_5$ | 0,886 | 0,892 | 0,886 | 0,869 | 0,856 | | | | | |
| | $I_6$ | 0,893 | 0,899 | 0,893 | 0,877 | 0,865 | 0,818 | | | | |
| | $I_7$ | 0,909 | 0,916 | 0,913 | 0,903 | 0,894 | 0,858 | 0,839 | | | |
| | $I_8$ | 0,921 | 0,929 | 0,929 | 0,926 | 0,921 | 0,896 | 0,881 | 0,856 | | |
| | $I_9$ | 0,924 | 0,933 | 0,935 | 0,938 | 0,937 | 0,921 | 0,909 | 0,888 | 0,864 | |
| | $I_{10}$ | 0,917 | 0,928 | 0,934 | 0,941 | 0,943 | 0,939 | 0,931 | 0,913 | 0,893 | 0,870 |
| | Average | 0,895 | 0,904 | 0,904 | 0,901 | 0,903 | 0,887 | 0,890 | 0,886 | 0,878 | 0,870 |

| | | 2004 subtraction of correlations (area minus interval) | | | | | | | | | |
|---|---|---|---|---|---|---|---|---|---|---|---|
| | | -0,002 | | | | | | | | | |
| | | 0,016 | 0,014 | | | | | | | | |
| | | 0,023 | 0,021 | 0,012 | | | | | | | |
| | | 0,022 | 0,020 | 0,011 | -0,015 | | | | | | |
| 2009 | | 0,022 | 0,019 | 0,015 | -0,006 | 0,009 | | | | | |
| | | 0,023 | 0,021 | 0,015 | -0,014 | 0,006 | 0,026 | | | | |
| | | 0,031 | 0,025 | 0,028 | 0,050 | 0,059 | 0,085 | 0,100 | | | |
| | | 0,013 | 0,005 | 0,007 | 0,042 | 0,047 | 0,067 | 0,079 | 0,100 | | |
| | | 0,015 | 0,006 | 0,006 | 0,031 | 0,035 | 0,048 | 0,057 | 0,074 | 0,099 | |
| | | 0,011 | -0,000 | -0,001 | 0,023 | 0,025 | 0,027 | 0,034 | 0,049 | 0,070 | 0,092 |
| | Sum | 0,173 | 0,131 | 0,092 | 0,112 | 0,179 | 0,252 | 0,270 | 0,223 | 0,169 | 0,092 |